\documentclass[iop]{emulateapj-rtx4} 
\shortauthors{Sekanina}
\shorttitle{Estimating Nuclear Dimensions of Comet C/1882 R1}
\slugcomment{Version \today }
\newcommand{\Rsun}{$R_{\mbox{\scriptsize \boldmath $\odot$}}\!$}

\begin{document}
\title{ESTIMATING DIMENSIONS OF THE NUCLEUS OF GREAT SEPTEMBER COMET OF 1882\\
       FROM MOTIONS OF ITS FRAGMENTS}
\author{Zdenek Sekanina}
\affil{Jet Propulsion Laboratory, California Institute of Technology,
  4800 Oak Grove Drive, Pasadena, CA 91109, U.S.A.}
\email{Zdenek.Sekanina@jpl.nasa.gov.}

\begin{abstract} 
Data on perihelion fragmentation of the Great September Comet of 1882 (C/1882~R1), a prominent
member of the Kreutz sungrazer system, are employed to estimate the size of the nucleus along
the radius vector at the time of splitting.  The prolate-spheroidal nucleus is assumed to
fragment tidally at perihelion along planes normal to this direction.  The relative velocities,
derived by Sekanina \& Chodas (2007) from revised positional-separation data on six
fragments collected originally by Kreutz (1888), are interpreted as measures of the
Sun's differential gravitational acceleration on the centers of mass of adjacent fragments
at the time of breakup and therefore a function of heliocentric distance.  Their total of
7.8~m~s$^{-1}$ is equivalent to nearly 38~km in the sum of distances of the centers of mass
along the radius vector and to the nuclear size of about 60~km.  The observed
sheath of diffuse material, the remains of the crumbling part of the nucleus that ended up
enveloping the train of the six fragments, included a population of less massive fragments.
They are expected to feed, over centuries in a distant future, an influx of dwarf
sungrazers reminiscent of the current stream of SOHO Kreutz comets.  It is speculated that
the parent comet of C/1882~R1 experienced similar fragmentation in the early 12th century
and one of its major fragments --- a bright sungrazer --- may return in the mid-21st century.
\end{abstract}

\section{Introduction}  
The brilliant sungrazer C/1882 R1, usually referred to as the Great September Comet of
1882, was exceptional in many ways.  One of the surprises was a major fragmentation event
that occurred at, or very close to, perihelion and turned the single, sizable nucleus
before perihelion into a string of fragments afterwards.  According to Kreutz (1888), the
elongation of the nuclear condensation was first noticed by O.\ de Bernardi\`eres, the
leader of the French Venus-transit expedition to Chile, on September~21, 1882, four days
after perihelion.  Barnard (1883) reported an elongated nucleus for the first time on
September~27.  On the other hand, in summarizing the monitoring of the comet at the
Cape Observatory, Gill (1911) stated that there was not ``any appearance of elongation
noted during the observations of September~28.''  On September~30, however, W.\,L.\,Elkin
saw at Cape an elongated nucleus with a heliometer~and~W.\,H.\,Finlay detected ``two
balls of light in the head'' with a 15-cm refractor (Gill 1883), a development also
noted by Kreutz (1888).  The nuclear region eventually developed into a chain of
nuclear condensations described as ``beads strung on a thread of worsted,''
except that the thread was itself buried in a sheath of diffuse material; the maximum
number of condensations was six, but usually reported were two or three.  The nuclear
train steadily growing in length with time and a parallel increase in the widths of the
gaps between the condensations were among the conspicuous properties of the feature.

The multiplicity of the nucleus and morphology of the fragments were extensively
monitored; Kreutz's (1888) compilation included almost 500~positional data (the sheath
dimensions, distances between condensations, and their orientations).  The most credible
data in Kreutz's collection were more recently reexamined by Sekanina \& Chodas (2007), who
applied Sekanina's (1982) model for split comets to correct errors and provide a table
of 102 well-documented offsets of five condensations --- A, C, D, E, and F --- from
condensation B (Kreutz's nucleus No.\,2), which was usually the brightest.  This exercise
resulted in establishing the magnitudes of relative velocities among the six condensations
at the time of fragmentation, the basis for the work described below.

\section{The Tidally-Driven Fragmentation of\\the Nucleus} 
Kreutz (1891) used available astrometric observations to compute separate sets of orbital
elements for four condensations, which showed a systematic increase in the orbital period
from 671~years for A (Kreutz's No.\,1) to 955~years for D (Kreutz's No.\,4).
With a constant perihelion distance, these values imply a difference in the orbital
velocity at perihelion of only 2.5~m~s$^{-1}$ between A and D.  Direct computation of
the relative velocity (Sekanina \& Chodas 2007) between the two fragments gives
5.6~m~s$^{-1}$.  This disparity implies that differences in the perihelion distance of
the centers of mass of the individual parts of the original nucleus at the moment they
became independent fragments due to the Sun's tidal forces were critical.

I now consider two point masses moving in very similar sungrazing orbits of the same
semimajor axis, $a$, connected with a massless rectilinear rigid thread of length
$\Delta \ell$, aligned with the radius vector.  Let the thread break at a heliocentric
distance $r_{\rm frg}$ at or close to perihelion, setting the two masses free to
continue moving, from that time on, in independent orbits with their own velocities
that at $r_{\rm frg}$ differ by $\Delta V$, solely because of effects of the Sun's
differential gravity resulting from the slightly uneven heliocentric distances,
$\Delta \ell$.  The relation between $\Delta \ell$ and $\Delta V$ is obtained by
differentiating the expression for the orbital velocity as a function of $r_{\rm frg}$,
\begin{equation}
\Delta \ell = -\frac{\sqrt{2}}{k_0} \, r_{\rm frg}^{\frac{3}{2}} \, \Delta \!\!\:V
 \,[ 1 \!+\! o(1)],  
\end{equation}
where $k_0$ is the Gaussian gravitational{\vspace{-0.03cm}} constant,~\mbox{equaling}
0.0172021\,AU$^{\frac{3}{2}}$day$^{-1}$\,and \mbox{$|o(1)| \!=\! {\textstyle \frac{1}{4}}
(r_{\rm frg}/a)$}.~The~\mbox{negative} sign means that the mass point closer to the
Sun acquires a slightly higher velocity and vice versa.  When{\vspace{-0.03cm}}
$r_{\rm frg}$ is not too far{\vspace{-0.04cm}} from the perihelion distance, $q$,
then \mbox{$|o(1)| \!<\! 10^{-4}$}.  When $r_{\rm frg}$ is expressed in {\Rsun}\,,
$\Delta V$ in{\vspace{-0.04cm}} m~s$^{-1}$, and $\Delta \ell$ in km, the constant
in Equation~(1) equals 2.25.\footnote{In the expression for $o(1)$, $r_{\rm frg}$
and $a$ are given in he same units.}  As an example,{\vspace{-0.03cm}} with
\mbox{$r_{\rm frg} = q = 1.67$ {\Rsun}}\, for C/1882~R1 and \mbox{$\Delta V = -1$
m s$^{-1}$}, one finds that \mbox{$\Delta \ell \simeq +5$ km}.

\section{Multiple Fragmentation of a Model Nucleus}  
The probability of tidal splitting of a comet's nucleus grows with its decreasing
heliocentric distance and increasing length along the radius vector.  In the following,
the presplit nucleus of C/1882~R1 is modeled as a prolate spheroid of uniform density
that fragments at a time when its long axis was directed at the Sun.  Although genuine
fragments are zigzag shaped, I assume that instead they separated from each other along
parallel planes that are perpendicular to the long axis and therefore to the radius vector.
The fragments are thus modeled as spheroidal segments; the ones nearest to, and farthest
from, the Sun, as spheroidal caps.
 
Let the semi-axes of the presplit spheroidal~\mbox{nucleus}~be
$\cal A$,\,$\cal B$,\,where~\mbox{${\cal B} \!<\! {\cal A}$}.~The~\mbox{coordinates}\,\mbox{$\{x,\,y,\,z\}$}\,of~a~point~on
the surface reckoned from the center satisfy the condition
\begin{equation}
\frac{x^2}{{\cal A}^2} + \frac{y^2 \! + z^2}{{\cal B}^2} = 1. 
\end{equation}
Let the plane perpendicular to the long axis (and the radius vector), along which a
fragment is assumed to have split off from the rest of the nucleus, be located at a
distance $x$ from the center of the spheroid.  The cross section of the fragment is
a circle{\vspace{-0.06cm}} whose radius is \mbox{$u = \sqrt{y^2 \!+ z^2} = {\cal B}
\sqrt{ 1 - x^2/{\cal A}^2}$}.  If $\rho$ is a constant density, the mass of the
spheroid's cap bounded by this plane at the one end and by the spheroid's pole,
whose \mbox{$x = {\cal A}$}, at the other end, is expressed as a fraction of the
spheroid's mass by
\begin{eqnarray}
\Im(\xi) & = & \frac{\displaystyle \rho \!\! \int_{x}^{\cal A} \!\!\! \pi u^2
 dx}{{\textstyle \frac{4}{3}}{\displaystyle \pi \rho {\cal A}{\cal B}^2}}
 = \frac{3}{2 {\cal A}^3} {\displaystyle \int_0}^{\xi} \!\! \left( {\cal A} \xi
 \!-\! {\textstyle \frac{1}{2}} \xi^2 \right) d \xi \nonumber \\[0.2cm]
 & = & \frac{\xi^2}{4 {\cal A}^2} \left( 3 \!-\! \frac{\xi}{\cal A} \right) \!, 
\end{eqnarray}
where $\xi$ is the cap's height, \mbox{$\xi = {\cal A} - x$}.  Let now $\cal D$ be the
maximum dimension of the presplit nucleus, \mbox{${\cal D} = 2 {\cal A}$}, and $\zeta$
a dimensionless quantity that measures the height of the spheroidal cap in units of
$\cal D$, \mbox{$\zeta = \xi/{\cal D}$} (\mbox{$0 \leq \zeta \leq 1$}).  As in Equation~(3),
$\Im(\zeta)$ is independent of the axial ratio ${\cal B}/{\cal A}$ and equals
\begin{equation}
\Im(\zeta) = 3 \, \zeta^2 \! \left( 1 - {\textstyle \frac{2}{3}} \zeta  \right). 
\end{equation}
\begin{figure}[t]
\vspace{-2.35cm}
\hspace{2.13cm}
\centerline{
\scalebox{0.67}{
\includegraphics{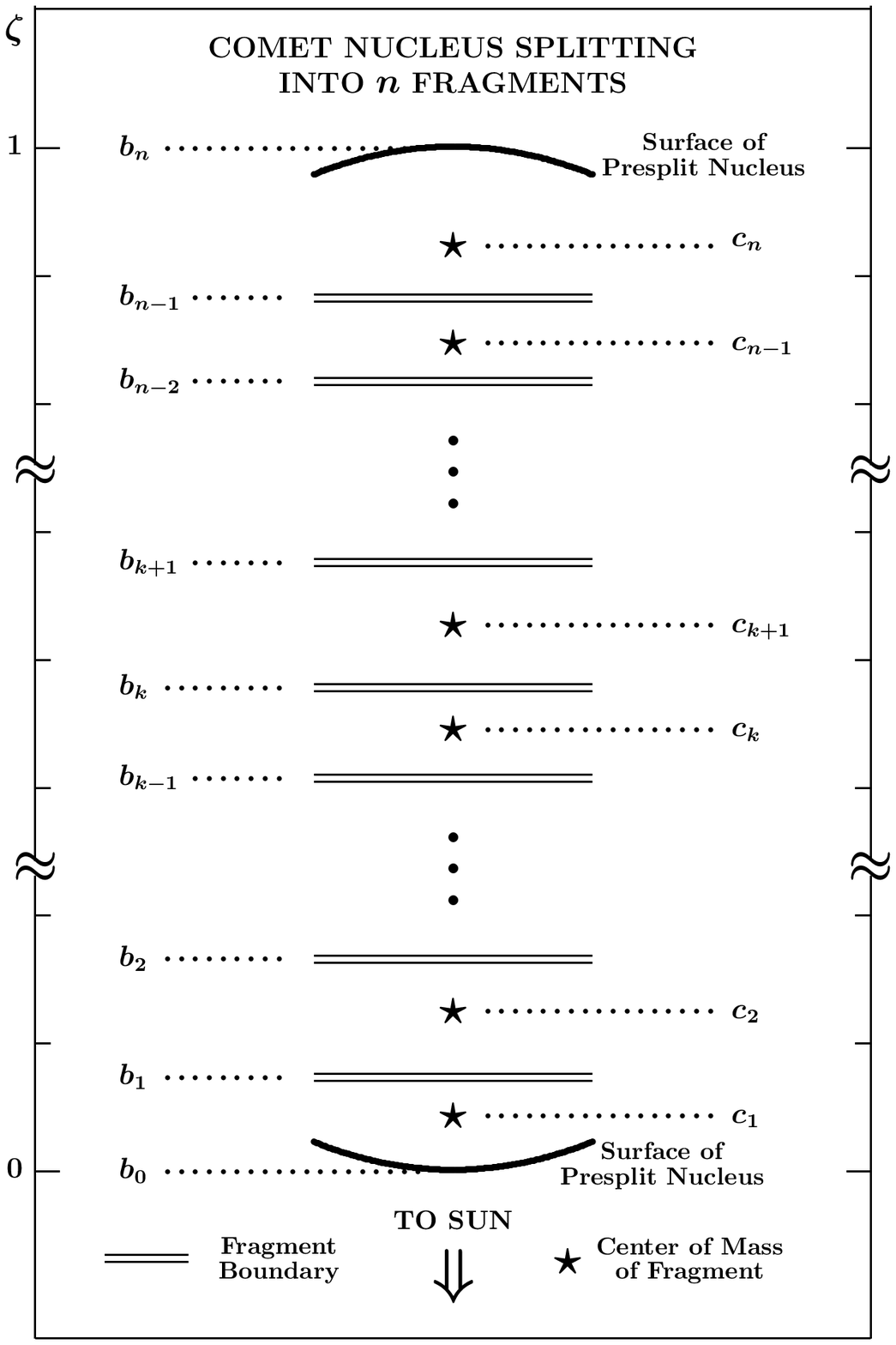}}}
\vspace{-4.75cm}
\caption{Schematic picture of the radial profile of a tidally split nucleus of normalized
dimensions:\ the sunward boundary is described by a coordinate \mbox{$\zeta = b_0 = 0$}, the
antisolar boundary by \mbox{$\zeta = b_n = 1$}.  The fragments' boundaries are marked by the
double lines, their centers of mass by stars. The distances between the mass centers of
adjacent fragments, $\Delta c_{k,k+1}$, are proportional to the relative velocities
$\Delta V_{k,k+1}$.  The contours of the presplit nucleus near the subsolar and antisolar
points are also depicted.{\vspace{0.4cm}}}
\end{figure}

\begin{table*}[t]
\vspace{-4.18cm}
\hspace{0.58cm}
\centerline{
\scalebox{1}{
\includegraphics{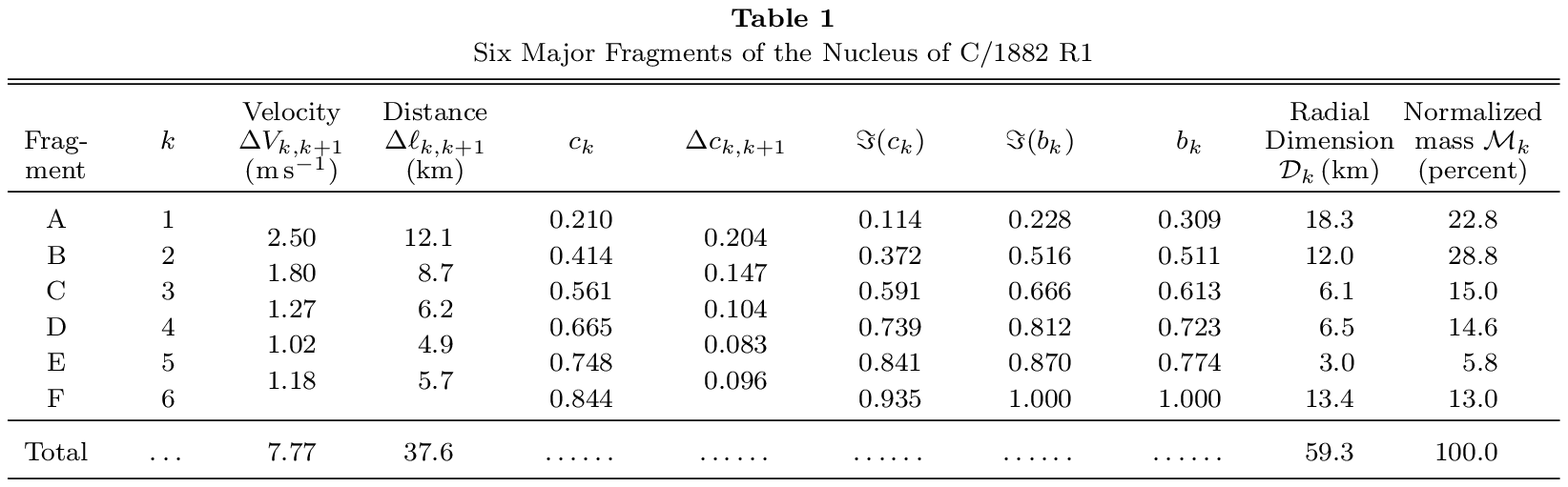}}}
\vspace{-19.8cm}
\end{table*}
It is noted that \mbox{$\Im(0) = 0$} and \mbox{$\Im(1) = 1$}.  For an arbitrary value
of $\Im(\zeta)$, $\zeta$ is obtained from the cubic equation,
\begin{equation}
\zeta = \frac{\sqrt{\Im(\zeta)}}{2 \cos\!\left\{ {\textstyle \frac{1}{3}} \!\arccos \left[
 -\sqrt{\Im(\zeta)}\, \right] \! \right\}}.  
\end{equation}

In the following I refer to $\zeta$ as the cap's normalized height and to $\Im(\zeta)$
as its normalized mass.  The normalized mass of a spheroidal segment is given as a
difference between the normalized masses of the respective spheroidal caps and a cap
can be considered a segment, one boundary plane of which degenerated into a point.
Figure~1 depicts a sequence of fragments that the pre\-split nucleus disintegrated into,
their total number being $n$.  The fragment nearest the Sun extends from the subsolar
point whose coordinate is \mbox{$\zeta = b_0 = 0$} to a boundary plane defined by the
coordinate \mbox{$\zeta = b_1 = 1$}.  A $k$-th fragment is confined, as Figure~1 shows, to
the segment bounded by the coordinates $b_{k-1}$ and $b_k$, so that its radial{\nopagebreak}
dimension amounts to
\begin{equation}
{\cal D}_k = {\cal D}(b_k - b_{k-1})  
\end{equation}
and its normalized mass is
\begin{equation}
{\cal M}_k = \Im(b_k) - \Im(b_{k-1}).  
\end{equation}

Next, the coordinate of the center of mass of the $k$-th fragment, $c_k$, is expressed
in terms of the coordinates of its boundary planes, $b_{k-1}$ and $b_k$, namely,
\begin{equation}
\Im(c_k) = {\textstyle \frac{1}{2}} \left[ \Im(b_{k-1}) + \Im(b_k) \right].  
\end{equation}
Specifically for $c_1$ one has \mbox{$\Im(c_1) = {\textstyle \frac{1}{2}} \Im(b_1)$}.

Equation (1) dictates the relationship of the relative velocity between the adjacent
fragments, $\Delta V_{k,k+1}$, and the difference in the radial distance, $\Delta
\ell_{k,k+1}$, which in turn is related to the difference between the normalized
distances of their centers of mass,
\begin{equation}
\Delta \ell_{k,k+1} = {\cal D} \, \Delta c_{k,k+1},  
\end{equation}
where
\begin{equation}
\Delta c_{k,k+1} = c_{k+1} - c_k.  
\end{equation}
Both the relative velocities $\Delta V_{k,k+1}$ and the distances $\Delta \ell_{k,k+1}$
add up to total the quantities in Equation~(1),
\begin{equation}
\Delta V = \sum_{k=1}^{n-1} \Delta V_{k,k+1}
\end{equation}
and
\begin{equation}
\Delta \ell = \sum_{k=1}^{n-1} \Delta \ell_{k,k+1}.  
\end{equation}
From Equations (9) through (12) it follows that
\begin{equation}
\Delta \ell = {\cal D}(c_n - c_1) 
\end{equation}
and because \mbox{$c_n - c_1 < 1$}, one always has \mbox{$\Delta \ell < {\cal D}$}.

\section{Numerical Solution and Results}
In the following I describe a solution for the nucleus of C/1882~R1 undergoing a
fragmentation event at perihelion, \mbox{$r_{\rm frg} = q = 1.67$ {\Rsun}}\,.  The
solution begins by guessing $c_1$ and $c_2$.  Because of Equation~(9) this also
means a guess for $\cal D$.  Next, $\Im(c_1)$ and $\Im(c_2)$ are derived from (4)
and, given that \mbox{$\Im(b_0) = 0$}, \mbox{$\Im(b_1) = 2 \Im(c_1)$} from (8) and
$b_1$ from (5).  From this point on, $c_{k+1}$ is derived from the input data via (9)
and (10), $\Im(c_{k+1})$ from (4), $\Im(b_{k+1})$ from (8), and $b_{k+1}$ again from (5),
until reaching $c_n$ and $b_n$.  If \mbox{$b_n \neq 1$}, the procedure is iterated
with a new pair of guessed values of $c_1$ and $c_2$.  Experimentation suggests that
$c_1$ should be increased when $b_n$ comes out to be smaller than unity, and vice
versa.  On the other hand, $c_2$ should be decreased relative to $c_1$ when, unacceptably,
\mbox{$\Im(b_{k+1}) < \Im(b_{k})$} for~any~$k$.

Turning now to Sekanina \& Chodas' (2007) paper, I note that the separation velocities
for fragments A, C, D, E, and F from B have magnitudes that are essentially independent
of whether they are taken to point in the radial or transverse direction.  Their averages
are taken to equal the velocities $\Delta V_{k,k+1}$ in Equation~(1).  The first four
columns of Table~1 list the input data.  Column~1 depicts the fragments' designations in
line with the paper by Sekanina \& Chodas, that is, A trailing (or nearest the Sun), F
leading (and seldom reported), and B usually the brightest and probably the most massive;
with the assigned numbers $k$ in column~2.  Column~3 shows the relative velocities B$-$A,
C$-$B, D$-$C, etc., averaged from the data in columns~8 and 10 of Table~7 in the paper
by Sekanina \& Chodas.  Column~4 presents the distances between the centers of mass of
the neighboring fragments computed from the relative velocities via Equation~(1) on the
premise that the fragmentation event occurred at perihelion, 1.67~{\Rsun} from the Sun.
This implies the minimum possible nuclear size ${\cal D}$.  The final iterated values of
the quantities derived to model the individual fragments are included in the remaining
columns, the primary ones in columns~5--9, the deduced ones in columns~10--11.

The size of the nucleus along the radius vector, the parameter that the method has been
conceived to determine, is found to reach a minimum of 59.3~km, when the fragmentation
event takes place at perihelion.  This size scales with the 1.5 power of the heliocentric
distance at fragmentation, so it would instead be 65~km if the comet breaks up 21~minutes
after (or before) perihelion; 80~km if 41~minutes off perihelion; and 100~km if 59~minutes
off perihelion (at which time the distance would be 2.36~{\Rsun}).  This end-to-end length
is very different from the length of the string of the fragments A--F, which, measured
by the derived positions of their centers of mass, is found to equal only 0.634 the total
length, or 37.6~km for perihelion fragmentation.  Comparison with the observations of the
dimensions of the sheath of diffuse material enveloping the six major fragments suggests
that when scaled to the distances between them, the length of the sheath did not exceed
the ratio of 1:0.634.  This indicates that the debris consisted of crumbled material of
the fragmenting nucleus that did not end up in one of the six major fragments.  The
particle size distribution of the debris is expected to have been limited mostly to
a range from many tens of meters to subkilometer-sized objects.

The model in Table 1 is in line with the assumption that fragment B was the most massive.
However, fragment A, not C, comes out to be the second most massive, while C as the third
most massive.  Yet, it is possible that the fragments at both ends, A and F, eventually
suffered more significant damage and a loss of mass than the rest and that their tabulated
masses are overestimates relative to the masses of the four inner fragments. 

\section{Predictions and Speculations} 
It is of interest to speculate about a spatial distribution of the debris as a function
of time.  As already mentioned briefly in Section~2, Kreutz (1891) determined the orbits
of the four nuclear condensations \mbox{A--D}, showing that their osculating heliocentric
orbital period, $P_k$, varied systematically from a minimum of 671~years for A (\mbox{$k
= 1$}) to a maximum of 955~years for D (\mbox{$k = 4$}).  A linear fit to $P_k$ as a
function of the normalized center-of-mass position parameter, $c_k$, of the four fragments
is given by
\begin{eqnarray}
P_k & = & 531 + 622 \, c_k,   \nonumber \\[-0.08cm]   
    &   & \!\!\pm23 \;\;\, \pm\!47
\end{eqnarray}
%
This expression fits the four fragments with a mean error of $\pm$16~years, which is
also in the range of typical deviations of a barycentric orbital period, required by
this problem, from its osculating equivalent.  Equation~(14) suggests that, apart from
nongravitational effects, the debris should have orbital periods near 530~years at
the sunward tip of the sheath and near 1150~years at its antisunward tip.  Returns to
perihelion should occur between the beginning of the 25th century and the first half
of the 31st century.  The fragments arriving over these six centuries should take the
appearance of SOHO-like dwarf comets.  At the rate of the current Kreutz SOHO sungrazers,
the total mass influx over the six centuries is estimated at less than 10$^{18}$\,grams
and possibly less than 10$^{17}$\,grams (Sekanina 2003).  Since the mass of a spherical
nucleus 60~km across with an assumed bulk density of 0.5~g~cm$^{-3}$ equals \mbox{$6
\times \! 10^{19}$}\,grams, the predicted arrival rate of these future SOHO-like dwarf
sungrazers is much higher than the rate of the currently arriving SOHO comets, even if
they contain only a modest fraction of the total mass.{\pagebreak}

The multiple breakup of C/1882~R1 raises the question of tidal
fragmentation of the comet's parent in the early 12th century.  Given the nearly
identical orbital elements of C/1882~R1 and C/1965~S1 (Ikeya-Seki), the result
of Marsden's (1967) integration of the accurately determined orbit of the main
nucleus of the latter comet back to its previous perihelion passage in AD~1116
provided a virtual proof that this comet and C/1882~R1 separated from their
common parent at about that time, although there are grounds to believe that
the parent was not the famous comet X/1106~C1 (Sekanina 2021).

Significantly, the difference of 83~years between the perihelion times of C/1965~S1
and C/1882~R1 is fairly close to a nearly constant difference of about 95~years
between the orbital periods of the adjacent nuclear fragments of C/1882~R1 (Kreutz
1891).  I recall that Strom (2002) noted that a sun-comet, a potential Kreutz object,
was recorded in a Chinese chronicle in the year 1792.  Furthermore, both Kreutz
(1901) and Marsden (1967) argued that the comet 1702a was a likely member of the
sungrazer system, moving in a path that fitted better the orbit of C/1882~R1 than
that of C/1843~D1.  With their presumed barycentric orbital periods near 580~years
and 670~years, respectively, the comets of 1702 and 1792 could be fragments of the
same parent as C/1882~R1 and C/1965~S1.  The average difference between the orbital
periods of the adjacent fragments among the foursome would be 88~years, and if their
12th century parent broke into at least five major parts, the arrival to perihelion
of yet another bright sungrazer, with an orbital period of about 930~years, is
expected around the year 2053, some three decades from now.\\[-0.2cm]

This research was carried out at the Jet Propulsion Laboratory, California Institute of
Technology, under contract with the National Aeronautics and Space Administration.\\[-0.2cm]

\begin{center}
{\footnotesize REFERENCES}
\end{center}

\vspace{-0.35cm}
\begin{description}
{\footnotesize
\item[\hspace{-0.3cm}]
Barnard, E. E. 1883, Astron. Nachr., 104, 267
\\[-0.57cm]
\item[\hspace{-0.3cm}]
Gill, D. 1883, Mon. Not. Roy. Astron. Soc., 43, 319
\\[-0.57cm]
\item[\hspace{-0.3cm}]
Gill, D. 1911, Ann. Cape Obs., 2, Pt. 1, 3
\\[-0.57cm]
%
\item[\hspace{-0.3cm}]
Kreutz, H. 1888, Publ. Sternw. Kiel, 3
\\[-0.57cm]
\item[\hspace{-0.3cm}]
Kreutz, H. 1891, Publ. Sternw. Kiel, 6
\\[-0.57cm]
\item[\hspace{-0.3cm}]
Kreutz, H. 1901, Astron. Abh., 1, 1
\\[-0.57cm]
\item[\hspace{-0.3cm}]
Marsden, B. G. 1967, AJ, 72, 1170
\\[-0.57cm]
\item[\hspace{-0.3cm}]
Sekanina, Z. 1982, in Comets, ed. L.\ L.\ Wilkening (Tucson, AZ:{\linebreak}
 {\hspace*{-0.6cm}}Univ. Arizona Press), 251
\\[-0.57cm]
\item[\hspace{-0.3cm}]
Sekanina, Z. 2003, ApJ, 597, 1237
\\[-0.57cm]
\item[\hspace{-0.3cm}]
Sekanina, Z. 2021, eprint arXiv 2109.01297
\\[-0.57cm]
\item[\hspace{-0.3cm}]
Sekanina, Z., \& Chodas, P. W. 2007, ApJ, 663, 657
\\[-0.65cm]
\item[\hspace{-0.3cm}]
Strom, R. 2002, A\&A, 387, L17}
\vspace{-0.39cm}
\end{description}
\end{document}